\documentclass[4apaper,numberedheadings,draft]{aipproc}
\layoutstyle{8x11single}
\usepackage{graphicx}
\usepackage{amsmath}
\usepackage{amsfonts}
\usepackage{amscd}

\usepackage{bm}

\begin{document}

%%% USER COMMANDS %%%%%%%%%%%%%%%%%%%%

% Dirac Ket
\newcommand{\Ket}[1]{\ensuremath{\left | #1 \right \rangle}}
% Dirac Bra
\newcommand{\Bra}[1]{\ensuremath{\left \langle #1 \right |}}
% Dirac Scalar Product
\newcommand{\BraKet}[2]{\ensuremath{\left \langle #1 \right |
\left. #2 \right \rangle}}
% Trace
\newcommand{\Tr}[1]{\ensuremath{\mbox{Tr} \left ( #1 \right )}}
% Partial Trace
\newcommand{\PTr}[2]{\ensuremath{\mbox{Tr}_{#1} \left ( #2 \right )}}

%%% FORMATTING COMMANDS %%%%%%%%%%%%%%%%%%%%

%%% TITLE PAGE %%%%%%%%%%%%%%%%%%%%

\title{Conditional Density Operators and the Subjectivity of Quantum Operations} 
\author{M. S. Leifer$^{1,2}$}{
address = {$^1$Perimeter Institute for Theoretical Physics, 31 Caroline Street North, Waterloo, Ontario, Canada, N2L 2Y5},
,altaddress = {$^2$Centre for Quantum Computing, Department of Applied Mathematics and Theoretical Physics, University of Cambridge, Wilberforce Road, Cambridge, CB3 0WA, UK}}

\date{November 17, 2006}

%%% ABSTRACT %%%%%%%%%%%%%%%%%%%%

\begin{abstract} 
Assuming that quantum states, including pure states, represent subjective degrees of belief rather than objective properties of systems, the question of what other elements of the quantum formalism must also be taken as subjective is addressed.  In particular, we ask this of the dynamical aspects of the formalism, such as Hamiltonians and unitary operators.  Whilst some operations, such as the update maps corresponding to a complete projective measurement, must be subjective, the situation is not so clear in other cases.  Here, it is argued that all trace preserving completely positive maps, including unitary operators, should be regarded as subjective, in the same sense as a classical conditional probability distribution.  The argument is based on a reworking of the Choi-Jamio{\l}kowski isomorphism in terms of ``conditional'' density operators and trace preserving completely positive maps, which mimics the relationship between conditional probabilities and stochastic maps in classical probability.  
\end{abstract}



\maketitle

%%% MAIN BODY %%%%%%%%%%%%%%%%%%%%

\section{Introduction}

In recent years, Caves, Fuchs and Shack (CFS) have argued that all quantum states, including pure states, should be taken to represent subjective degrees of belief rather than objective properties of systems \cite{CFSBayesian, CavFucSha02, Fuchs02, Fuchs03, CFS06}, in close analogy to the radical probabilist view of classical probabilities \cite{Jeffrey, deFinetti}.  The purpose of this article is not to debate the merits of this view, which have been extensively discussed elsewhere \cite{Hagar1, Hagar2, FuchsSam1, FuchsSam2, FuchsSam3}, but rather to investigate its consequences for the rest of the quantum formalism.  In particular, we address the question of whether quantum dynamics, variously expressed as Hamiltonians, unitary operators and Trace Preserving Completely Positive (TPCP) maps, should also be taken to represent subjective degrees of belief.  We argue that this is the case for \emph{all} TPCP maps, in the same sense that all conditional probabilities are subjective in the radical probabilist view of classical probability.

CFS have already argued that \emph{some} CP maps must be taken as subjective \cite{CFS06}.  For example, consider a non-destructive measurement in an orthonormal basis.  In the orthodox approach to quantum theory, on obtaining an outcome corresponding to a pure state $\Ket{\psi}$, the state of the system is updated to $\Ket{\psi}$, regardless of the initial state of the system.  However, for CFS there can never be a situation in which two agents are compelled to assign the same state to a system, even if they have access to exactly the same data.  As in radical probabilism, provided the two agents start with distinct enough prior beliefs, they need never converge on a common set of beliefs, regardless of how much data they share\footnote{Of course, in practical situations it is often reasonable to assume that the agents don't hold such singular beliefs, and then their views can be expected to converge.  Nevertheless, as a point of principle, incompatible beliefs are not labeled as irrational \emph{a priori}.}.   Thus, the projector $\Ket{\psi}\Bra{\psi}$ corresponding to the measurement outcome must be subjective, depending as it does on an analysis of the workings of the measurement device, and this analysis may differ for the two agents.  There are also clearly situations in which the subjectivity of quantum states can infect quantum operations.  For example, suppose that two agents both agree on the unitary evolution that applies to a joint system composed of a system of interest and its environment.  They will generally use different dynamical maps to describe the evolution of the system of interest alone, by virtue of the fact that that they may assign different initial states to the environment.  

The situation is less clear when considering a unitary operation on the system of interest alone, since in this case the environmental state is irrelevant to the action of the operation on the system, and unitary operations do not cause convergence of distinct states.  Thus, unlike the previously discussed cases, the subjectivity of unitary operations cannot be argued directly from the subjectivity of quantum states.  Therefore, it would not be inconsistent for CFS to hold onto the objectivity of unitary operations, which might be tempting, since the specification of a Hamiltonian seems to encode the objective content of our most successful physical laws.  However, Fuchs has rejected this road, and argues that all TPCP maps, including unitary operations, are analogous to conditional probabilities and so they should be taken to represent subjective degrees of belief \cite{Fuchs02, Fuchs03, FuchsSam2}.  This view could also be seen as implicit in the de Finetti theorem for quantum operations \cite{ProcDeF1, ProcDef2}, which does not single out unitary operations for any special treatment. 

Here, we significantly strengthen the case for the subjectivity of all TPCP maps by demonstrating a thoroughgoing analogy between TPCP maps and conditional probabilities, of the type needed to make Fuchs' arguments compelling to adherents of the CFS view.  In fact, we argue for a reconsideration of the domain of applicability of bipartite quantum states themselves.  Instead of assuming that they are always descriptions of a pair of distinct systems, we argue that they can also be used to describe the same system at two distinct instances of time in a ``prepare and measure'' scenario.  That this can be done in special cases has been known for quite a while in the context of quantum cryptography \cite{BenBraMer92}, where prepare and measure schemes are regularly traded for entanglement based schemes in proofs of the security of quantum key distribution \cite{ShorPreskill}, and this correspondence is generalized here.  There does not seem to be any substantive difference in the role played by the bipartite state in the two scenarios, so we argue that if the state is taken to be subjective in one context, then it should also be subjective in the other.  From this, the subjectivity of all quantum operations may be inferred. No doubt, this conclusion will seem unappealing to many hard-nosed physicists.   If unitary operations cannot be taken as objective then neither can Hamiltonians, and it seems that we may be in danger of losing the objectivity of physical laws altogether.  I argue that this fear is unfounded and rests on the same sort of category error as the identification of certainty with a subjective probability equal to one \cite{CFS06}.

From a broader perspective, this work suggests that it may be possible to cleanly separate the probabilistic and statistical parts of the quantum formalism from those that depend on its particular physical realization.  Despite the fact that the abstract formalism of quantum theory looks like a noncommutative generalization of classical probability, it still does not quite achieve a full separation, because it is necessary to know whether two events refer to distinct physical systems or to the same system at two different times in order to know how to combine them, i.e. whether to use the tensor product or a dynamical map.  In this respect, quantum theory is in closer analogy to the theory of stochastic process \cite{Doob42} than it is to abstract Kolmogorov probability theory \cite{Kolmogorov29}, since the latter is independent of any identification of events in an abstract sample space with physical events in spacetime.  We believe that a more Kolmogorovian formulation of quantum theory would offer new insights into quantum information protocols, as well as clarifying foundational issues, and regard the current work as a step towards such a formalism. 

The remainder of this paper is structured as follows. In section \ref{Prelim} the finite dimensional $C^*$-algebraic formalism of quantum theory is briefly reviewed.  In section \ref{CDO}, the ``conditional density operator'' is introduced, which is the main tool for relating bipartite quantum states to TPCP maps.  In section \ref{CJI}, the Choi-Jamio{\l}kowski isomorphism is discussed.  Whilst this is well-known, we give a novel presentation in which the isomorphism is taken to relate \emph{conditional} density operators to TPCP maps, rather than relating unnormalized bipartite states to general CP maps as in usual presentations \cite{Jam72, Choi75, VerVer03, ArrPat04, ZyczBeng04, Griffiths05}. Section \ref{CJI:Trad} gives the traditional operational interpretation of the isomorphism in terns of noisy gate teleportation and section \ref{CJI:Stoch} gives a different operational interpretation by which the analogy to the role of conditional probability in classical stochastic processes is made clear.  In section \ref{Subjective}, the argument for the subjectivity of quantum operations is given and in \S\ref{Reconcile}, we argue that the subjectivity of quantum operations does not imply the subjectivity of physical laws.  Finally, section \ref{Conc} contains a summary and conclusions. 

The technical results presented here generalize those of previously published work \cite{Leifer06} from the finite dimensional Hilbert space formalism to finite dimensional $C^*$ algebras.  The current treatment places greater emphasis on the role of the conditional density operator, which we hope clarifies the physical interpretation given in \cite{Leifer06}.

\section{Preliminaries}

\label{Prelim}

For present purposes it is convenient to work in the $C^*$-algebraic formalism for quantum theory.  This facilitates the comparison between classical probability and quantum theory, since the former is obtained whenever the algebra is commutative.  Because we are concerned mainly with conceptual matters, it is convenient to specialize to finite dimensional algebras in order to avoid analytical complications.  Any such algebra can be thought of as the algebra of block-diagonal matrices on a finite dimensional Hilbert space, once a basis for the latter is fixed.  Hence, the most general algebra we are concerned with is
\begin{equation}
\mathfrak{A} = \mathfrak{B}(\mathbb{C}^{d_1}) \oplus \mathfrak{B}(\mathbb{C}^{d_2}) \oplus \ldots \oplus \mathfrak{B}(\mathbb{C}^{d_n}),
\end{equation}
where $\mathfrak{B}(\mathcal{H})$ is the algebra of bounded operators on a Hilbert space $\mathcal{H}$.

Two important special cases are the classical commutative algebras, $\mathfrak{B}(\mathbb{C})^{\oplus n}$, which are diagonal, and the irreducible ``full quantum'' algebras, $\mathfrak{B}(\mathbb{C}^d)$.  States on an algebra are usually defined as positive linear functionals $\omega: \mathfrak{A} \rightarrow \mathbb{C}$ that satisfy $\omega(I) = 1$, where $I$ is the identity operator in $\mathfrak{A}$.  In the finite dimensional case, these can be replaced by density matrices $\rho \in \mathfrak{A}$ that are positive and have unit trace via the identification $\omega(A) = \Tr{A\rho}$ for all $A \in \mathfrak{A}$.  Given two independent subsystems corresponding to algebras $\mathfrak{A}_A$ and $\mathfrak{A}_B$, the combined system corresponds to the tensor product $\mathfrak{A}_A \otimes \mathfrak{A}_B$, which coincides with a Cartesian product of sample spaces in the case where both algebras are classical, and the usual tensor product of Hilbert spaces when both algebras are irreducible.  Given a state on the tensor product $\rho_{AB} \in \mathfrak{A}_A \otimes \mathfrak{A}_B$, the reduced states $\rho_A \in \mathfrak{A}_A$ and $\rho_B \in \mathfrak{A}_B$ are given by $\rho_A = \PTr{B}{\rho_{AB}}$ and $\rho_B = \PTr{A}{\rho_{AB}}$.

The most general dynamics of a system is given by a linear map $\mathcal{E}_{B|A} :\mathfrak{A}_A \rightarrow \mathfrak{A}_B$, where the input and output systems are generally allowed to be different.  Here, this is taken to be a map acting on density operators, i.e. we are working in a Schr{\"o}dinger picture, which is unproblematic in finite dimensions.  The map should be \emph{Completely Positive} (CP), meaning that $\mathcal{E}_{B|A} \otimes \mathcal{I}_C:\mathfrak{A}_A \otimes \mathfrak{A}_C \rightarrow \mathfrak{A}_B \otimes \mathfrak{A}_C$ is a positive map for all finite dimensional algebras $\mathfrak{A}_C$, and where $\mathcal{I}_C:\mathfrak{A}_C \rightarrow \mathfrak{A}_C$ is the identity map on $\mathfrak{A}_C$.  Furthermore, if no measurements are performed then the map should be \emph{Trace Preserving} (TP) in order to maintain the normalization of density operators. 

\section{Conditional Density Operator}

\label{CDO}

In classical probability, the conditional probability of an event $Y$, given an event $X$ is defined as
\begin{equation}
P(Y|X) = \frac{P(X \cap Y)}{P(X)},
\end{equation}
wherever $P(X) \neq 0$ and is undefined otherwise.  Defining an analog of this for general $C^*$-algebraic theories is a tricky problem, and there are several alternative possibilities.  Here, we only deal with a special case, which is however the most important for practical applications.  Consider a tensor product of two classical algebras $\mathfrak{A}_A \otimes \mathfrak{A}_B$ with corresponding bases $\{\Ket{j}_A\}, \{\Ket{k}_B\}$ in which the operators are diagonal.  A state $\rho_{AB}$ on this algebra can be written in terms of its diagonal components $ (\rho_{AB})_{jk,jk}$ as 
\begin{equation}
\rho_{AB} = \sum_{jk} (\rho_{AB})_{jk,jk} \Ket{j}\Bra{j}_A \otimes \Ket{k}\Bra{k}_B,
\end{equation}
and the reduced state on system $A$ is given by $\rho_A = \PTr{B}{\rho_{AB}}$, with diagonal components $(\rho_A)_{jj} = \sum_k (\rho_{AB})_{jk,jk}$.  Now, the conditional probability that system $B$ is in state $\Ket{k}_B$, given that system $A$ is in state $\Ket{j}_A$ is given by $\frac{(\rho_{AB})_{jk,jk}}{(\rho_{A})_{j,j}}$, provided $(\rho_{A})_{j,j}$ is nonzero.  This can be written as a matrix of conditional probabilities, given by
\begin{equation}
(\rho_{B|A})_{jk,jk} = \frac{(\rho_{AB})_{jk,jk}}{(\rho_{A})_{j,j}}, 
\end{equation}
or in operator notation
\begin{equation}
\label{CDO:ClassOp}
\rho_{B|A} = \left ( \rho_A^{-1} \otimes I_B \right ) \rho_{AB},
\end{equation}
where $I_B$ is the identity operator in $\mathfrak{A}_B$.
Here, care must be taken when $\rho_A$ is not of full rank, in which case we may restrict the domain of $\rho_{B|A}$ to the support of $\rho_A^{-1} \otimes I_B$.  An alternative is to define the generalized inverse of $\rho_A$ to have the same eigenspaces as $\rho_A$, with eigenvalue zero on the null eigenspace of $\rho_A$ and reciprocal eigenvalues on all other eigenspaces.  This is the approach we adopt throughout.

In the general noncommutative case, it should be clear that eq. (\ref{CDO:ClassOp}) can be generalized in many different ways, due to the fact that $\rho_A^{-1}\otimes I_B$ and $\rho_{AB}$ need not commute.  In doing so, one should bear in mind the various possible applications of conditional probability (e.g. the updating of probabilities by Bayesian conditionalization, in stochastic processes, and in information theory) and check that the chosen generalization is useful for describing sensible quantum analogs of at least some of these.  The alternative, to focus on formal mathematical properties of conditional probability, may also be a useful approach, but is unlikely to lead to applicable concepts on its own.  In this regard, the following generalization suggests itself as particularly interesting\footnote{As pointed out by Cerf and Adami \cite{CerfAdami97, CerfAdami98, CerfAdami99}, another definition of note is $\rho_{B|A} = \lim_{n \rightarrow \infty} \left ( \rho_A^{-\frac{1}{2n}} \otimes I_B \rho_{AB}^{\frac{1}{n}} \rho_A^{-\frac{1}{2n}} \otimes I_B \right )^n$, since this allows the von Neumann conditional entropy to be expressed as $S(B|A) = - \Tr{\rho_{AB} \log \rho_{B|A}}$ in analogy to the classical expression for conditional Shannon entropy.}:
\begin{equation}
\label{CDO:Define}
\rho_{B|A} = \left ( \rho_A^{-\frac{1}{2}} \otimes I_B \right ) \rho_{AB} \left ( \rho_A^{-\frac{1}{2}} \otimes I_B \right ).
\end{equation}
This equation may be inverted to obtain
\begin{equation}
\label{CDO:Reverse}
\rho_{AB} = \left ( \rho_A^{\frac{1}{2}} \otimes I_B \right ) \rho_{B|A} \left ( \rho_A^{\frac{1}{2}} \otimes I_B \right ).
\end{equation}
Note that $\rho_{B|A}$ is a positive operator, since it is of the form $A^\dagger A$ for $A = \rho_A^{-\frac{1}{2}} \otimes I_B \rho_{AB}^{\frac{1}{2}}$, but is not a density operator because it does not have unit trace.  In fact, $\PTr{B}{\rho_{B|A}} = I_{\text{supp}(\rho_A)}$, where $I_{\text{supp}(\rho_A)}$ is the projector onto the support of $\rho_A$, so the trace of $\rho_{B|A}$ is the rank of $\rho_A$.  In the classical case, this corresponds to the fact that the matrix of conditional probabilities $(\rho_{B|A})_{jk,jk}$ must give a valid probability distribution for each value of $j$, i.e. $\sum_{k}(\rho_{B|A})_{jk,jk} = 1$.

In line with the earlier warning, it should be checked that this definition of a quantum conditional density operator actually plays a role in applications.  In \S\ref{CJI:Stoch}, it is shown that the relation between conditional density operators and TPCP maps is analogous to the relation between conditional probabilities and stochastic matrices in a classical stochastic process.  The conditional density operator is also related to Fuchs' proposal for a quantum analog of Bayesian conditionalization \cite{Fuchs02, Fuchs03}, and the analog of Bayes' rule, $\rho_{B|A} = \rho_A^{-\frac{1}{2}} \otimes \rho_B^{\frac{1}{2}} \rho_{A|B} \rho_A^{-\frac{1}{2}} \otimes \rho_B^{\frac{1}{2}}$, is relevant to the problem of pooling quantum states, both of which are described in forthcoming work \cite{LeiferSpekkens}.

\section{The Choi-Jamio{\l}kowski Isomorphism}
\label{CJI}

The central tool used in the arguments below is the isomorphism discovered by Jamio{\l}kowski \cite{Jam72}, and developed by Choi \cite{Choi75}, between Completely Positive maps $\mathfrak{A}_A \rightarrow \mathfrak{A}_B$ and (generally unnormalized) states in $\mathfrak{A}_A \otimes \mathfrak{A}_B$\footnote{Jamio{\l}kowski and Choi both take $\mathfrak{A}_A =\mathfrak{B}(\mathbb{C}^d)$, but the extension to general finite dimensional algebras is straightforward as shown below}.  For present purposes, it is convenient to formulate it as an isomorphism between \emph{Trace-Preserving} Completely Positive maps $\mathfrak{A}_A \rightarrow \mathfrak{A}_B$ and \emph{conditional} density operators in $\mathfrak{A}_A \otimes \mathfrak{A}_B$.  This formulation gives greater intuition about the physical meaning of the isomorphism, as shown in \S\ref{CJI:Stoch}.  

We begin with the case where $\mathfrak{A}_A = \mathfrak{B}(\mathbb{C}^{d_A})$ and $\mathfrak{A}_B$ is a general finite dimensional algebra, and then generalize to the case of general finite dimensional $\mathfrak{A}_A$ below.  Let $\mathcal{E}_{B|A}:\mathfrak{A}_A \rightarrow \mathfrak{A}_B$ be a TPCP map.  To define the isomorphism, we begin with the $\mathcal{E}_{B|A} \rightarrow \rho_{B|A}$ direction.  Let $\mathfrak{A}_{A'}$ be another copy of the algebra $\mathfrak{A}_A$, i.e. $\mathfrak{A}_{A'} = \mathfrak{A}_A = \mathfrak{B}(\mathbb{C}^{d_A})$.  The isomorphism is dependent on an arbitrary choice of basis for $\mathbb{C}^{d_A}$, so let $\{\Ket{j}_A\}$ be such a basis and define the ``maximally entangled'' conditional state vector on $\mathbb{C}^{d_A} \otimes \mathbb{C}^{d_A}$ as
\begin{equation}
\Ket{\Phi^+}_{A'|A} = \sum_{j=1}^{d_A} \Ket{jj}_{A'A}.
\end{equation}
This is so called because when one uses eq. (\ref{CDO:Reverse}) to combine the conditional state $\rho^+_{A'|A} = \Ket{\Phi^+}_{A'|A}\Bra{\Phi^+}_{A'|A}$ with a maximally mixed marginal state $\rho_A = \frac{I_A}{d_A}$, where $I_A$ is the identity operator in $\mathfrak{A}_A$, one obtains a properly normalized maximally entangled state $\rho^+_{AA'} = \Ket{\Phi^+}_{AA'}\Bra{\Phi^+}_{AA'}$, where $\Ket{\Phi^+}_{AA'} = \frac{1}{\sqrt{d_A}} \Ket{\Phi^+}_{A'|A}$.  However, note that $\rho^+_{A'|A}$ generally does not yield a maximally entangled state when combined with an arbitrary reduced state $\rho_A$.

Next, we define the conditional state $\rho_{B|A}$ associated with the map $\mathcal{E}_{B|A}$ via
\begin{equation}
\label{Iso:Forward}
\rho_{B|A} = \mathcal{E}_{B|A'} \otimes \mathcal{I}_A \left ( \rho^+_{A'|A} \right ),
\end{equation}
where $\mathcal{I}_A$ is the identity CP-map on system $A$.  Note that here $\mathcal{E}_{B|A'}$ is acting on the ancillary system $A'$, transforming it into system $B$.  It is straightforward to check that $\rho_{B|A}$ is a valid conditional state, which is transformed into a valid joint state $\rho_{AB}$ when it is combined with any reduced density operator $\rho_A$ in $\mathfrak{A}_A$ via eq. (\ref{CDO:Reverse}).

For the $\rho_{B|A} \rightarrow \mathcal{E}_{B|A}$ direction, note that the action of $\mathcal{E}_{B|A}$ on an arbitrary state $\sigma_A \in \mathfrak{A}_A$ may be recovered from $\rho_{B|A}$ via
\begin{equation}
\label{Iso:Back}
\mathcal{E}_{B|A} \left ( \sigma_A \right ) = \PTr{AA'}{ \left ( \rho^+_{A'|A} \otimes I_B \right ) \left ( \sigma_A \otimes \rho_{B|A'} \right )},
\end{equation}
which is easily checked by expanding the states in the basis used to define the isomorphism.  Note that the state $\rho_{B|A}$ is pure iff the TPCP map $\mathcal{E}_{B|A}$ is an isometry, and in the case where $\mathfrak{A}_A = \mathfrak{A}_B$, this means that $\mathcal{E}_{B|A}$ is unitary. 

Finally, we briefly explain how to extend the isomorphism to the case where $\mathfrak{A}_A$ is an arbitrary finite dimensional algebra.  The problem is that, for a general algebra of the form  $\mathfrak{A}_A = \mathfrak{B}(\mathbb{C}^{d_1}) \oplus \mathfrak{B}(\mathbb{C}^{d_2}) \oplus \ldots \oplus \mathfrak{B}(\mathbb{C}^{d_n})$, $\mathfrak{A}_A \otimes \mathfrak{A}_{A'}$ does not contain the conditional state $\rho^{+}_{A'|A}$, since $\Ket{\Phi^+}_{A'|A}$ is a superposition of all basis states of the form $\Ket{jj}_{AA'}$, and this is ruled out for any algebra which is the direct sum of more than one irreducible component.  To resolve this, note that $\mathfrak{A}_A$ may be embedded in $\mathfrak{B}(\mathbb{C}^{d_1+d_2+\ldots + d_n})$ by associating operators in $\mathfrak{A}_A$ with block-diagonal matrices in $\mathfrak{B}(\mathbb{C}^{d_1+d_2+\ldots + d_n})$.  The action of $\mathcal{E}_{B|A}$ is not well defined on this algebra, since its domain is $\mathfrak{A}_A$, but this can be dealt with by introducing the projection map $\mathcal{P}:\mathfrak{B}(\mathbb{C}^{d_1+d_2+\ldots + d_n}) \rightarrow \mathfrak{A}_A$, which can be written in the form
\begin{equation}
\label{Iso:Project}
\mathcal{P}(\rho) = \sum_{j=1}^n P_j \rho P_j,
\end{equation}
where $P_j$ is the projector onto the factor $\mathbb{C}^{d_j}$ in $\mathbb{C}^{d_1+d_2+\ldots + d_n}$.  Now, $\mathcal{E}_{B|A'}$ may be replaced with $\tilde{\mathcal{E}}_{B|A'} = \mathcal{E}_{B|A'}\circ \mathcal{P}_{A'}$ in eq. (\ref{Iso:Forward}), and this map is well defined on $\mathfrak{B}(\mathbb{C}^{d_1+d_2+\ldots + d_n})$.  Since $\mathcal{P}$ is idempotent, one may additionally replace $\rho^+_{A'|A}$ with 
\begin{equation}
\tilde{\rho}^+_{A'|A} = \mathcal{I}_A \otimes \mathcal{P}_{A'} (\rho_{A'|A})
\end{equation}
in eqs. (\ref{Iso:Forward}) and (\ref{Iso:Back}), which is a well-defined conditional state in $\mathfrak{A}_A \otimes \mathfrak{A}_{A'}$.  The actions of $\tilde{\mathcal{E}}_{B|A'}$ and $\mathcal{E}_{B|A'}$ on this state are identical, so we obtain 
\begin{equation}
\label{Iso:Forward2}
\rho_{B|A} = \mathcal{E}_{B|A'} \otimes \mathcal{I}_A \left ( \tilde{\rho}^+_{A'|A} \right )
\end{equation}
and
\begin{equation}
\label{Iso:Back2}
\mathcal{E}_{B|A} \left ( \sigma_A \right ) = \PTr{AA'}{\tilde{\rho}^+_{A'|A} \otimes I_B \sigma_A \otimes \rho_{B|A'}},
\end{equation}
as the generalized version of the isomorphism.

\subsection{Operational Interpretation in terms of teleportation}

\label{CJI:Trad}

There is a standard interpretation of the Choi-Jamio{\l}kowski isomorphism in terms of ``noisy gate teleportation'', which is the generalization of a protocol considered in \cite{NielChuang97} from unitary operations to arbitrary TPCP maps. To describe this, we begin with the case where $\mathfrak{A}_A = \mathfrak{B}(\mathbb{C}^{d_A})$, and combine the conditional states, $\rho_{B|A'}$ and $\rho^+_{A'|A}$, with maximally mixed reduced states, $\rho_{A'} = \frac{I_{A'}}{d_A}$ and $\rho^+_{A} = \frac{I_{A}}{d_{A}}$, via eq. (\ref{CDO:Reverse}), so that the reverse direction of the isomorphism eq. (\ref{Iso:Back}) can be rewritten in terms of the properly normalized joint states $\rho_{A'B} = \frac{1}{d_A} \rho_{B|A'}$ and $\rho^+_{AA'} = \frac{1}{d_A} \rho_{A'|A}$ as
\begin{equation}
\label{CJI:Telep}
\mathcal{E}_{B|A}(\sigma_A) = d_A^2 \PTr{AA'}{\rho^+_{AA'} \otimes I_B \sigma_A \otimes \rho_{A'B}}.
\end{equation}
Now, suppose that Alice holds a system in an unknown state\footnote{For the subjectivist, the phrase ``unknown state'' should set alarm bells ringing.  It is a shorthand for saying that the system is prepared by Charlie, who then gives it to Alice without revealing any details of the preparation procedure.  The ``unknown state'' is the one assigned by Charlie \cite{CavFucSha02}.} $\sigma_{A} \in \mathfrak{A}_A$ and that Alice and Bob share a pair of systems in the state $\rho_{A'B}$.  They would like for Bob to end up with his system in the transformed state\footnote{Again, it is Charlie's description of Bob's state that is being referred to.} $\mathcal{E}_{B|A} \left ( \sigma_A \right )$, using only local operations and classical communication and the state $\rho_{A'B}$ as resources.  To achieve this, Alice can make a joint measurement of the systems $A$ and $A'$ in a basis that includes the state $\rho_{AA}^+$.  If the outcome corresponding to this state is obtained, then the procedure is successful, which may be deduced from eq. (\ref{CJI:Telep}).  It is also evident from eq. (\ref{CJI:Telep}) that the probability of obtaining this successful outcome is $\frac{1}{d_A^2}$.  On the other hand, if the $\rho^+_{AA'}$ outcome is not obtained then the procedure fails.  In some cases it is still possible for Bob to reconstruct the state $\mathcal{E}_{B|A}(\sigma_A)$  by applying a local operation that depends on Alice's outcome, which she can inform him of via classical communication.  In particular, this happens when $\mathcal{E}_{B|A}$ is the identity, in which case we obtain the standard teleportation protocol \cite{Teleport}.

This protocol can be straightforwardly generalized to the case where $\mathfrak{A}_A$  is a general finite dimensional algebra.  However, the expression for the success probability becomes more complicated because more than one state may be mapped to $\tilde{\rho}^+_{A'|A}$ by the action of $\mathcal{P}_{A'}$.  In particular, it can happen that states associated failure outcomes in $\mathfrak{B}(\mathbb{C}^{d_1 + d_2 + \ldots + d_n})$ are mapped to the success outcome $\tilde{\rho}^+_{A'|A}$ by $\mathcal{P}_{A'}$, which increases the probability of success.  As an example, consider the classical algebra $\mathfrak{B}(\mathbb{C}) \oplus \mathfrak{B}(\mathbb{C})$ and its embedding in $\mathfrak{B}(\mathbb{C}^2)$.  Here, Alice should make a measurement in the Bell basis
\begin{equation}
\begin{array}{ll}
\Ket{\Phi^+} = \frac{1}{\sqrt{2}} \left ( \Ket{00} + \Ket{11} \right ) & \Ket{\Psi^+} = \frac{1}{\sqrt{2}} \left ( \Ket{01} + \Ket{10} \right ) \\
\Ket{\Phi^-} = \frac{1}{\sqrt{2}} \left ( \Ket{00} - \Ket{11} \right )  & \Ket{\Psi^-} = \frac{1}{\sqrt{2}} \left ( \Ket{01} - \Ket{10} \right ),
\end{array}
\end{equation}
as in the teleportation protocol.  Under the projection map $\mathcal{P}_{A'}$,
\begin{equation}
\rho^+_{AA'} = \Ket{\Phi^+}\Bra{\Phi^+}_{AA'} \rightarrow \tilde{\rho}^+_{AA'} = \frac{1}{2} \left ( \Ket{00}\Bra{00} + \Ket{11}\Bra{11} \right )_{AA'},
\end{equation}
but $\Ket{\Phi^-}\Bra{\Phi^-}_{AA'}$ also gets mapped to the same thing, so these outcomes may be grouped together and the success probability is increased from $1/4$ to $1/2$.  Similarly, the failure outcomes $\Ket{\Psi^\pm}\Bra{\Psi^\pm}_{AA'}$ both get mapped to $\frac{1}{2} \left ( \Ket{01}\Bra{01} + \Ket{10}\Bra{10} \right )_{AA'}$, so these may also be grouped together and Alice's measurement is then just a parity measurement of her two classical bits.  In the case where $\mathcal{E}_{B|A}$ is the identity, Bob can recover the correct state by flipping his bit when Alice gets the failure outcome and the whole procedure is simply a classical one-time-pad (Vernam cipher) \cite{Vernam, ShannonCrypt}.  The similarity between teleportation and the one-time pad has been remarked upon before \cite{KonTelep, CollinsPop}, but in the algebraic formulation it is more than just a similarity.  They both arise from the same isomorphism, so they are, in fact, the same thing.

\subsection{Operational interpretation in terms of stochastic processes}

\label{CJI:Stoch}

The previous interpretation resulted from combining the conditional states with maximally mixed reduced states, so it is natural to ask whether there is any interpretation that results from combining $\rho_{B|A}$ with an arbitrary reduced state $\rho_A$.  Doing so reveals the Choi Jamio{\l}kowski isomorphism to be a generalization of the relationship between stochastic dynamics and conditional probabilities in classical probability theory.  

In the classical case, it is a familiar fact that we can always describe the correlations between two random variables by a joint probability distribution, regardless of whether the variables refer to two distinct physical systems or to the same quantity associated with the same system at two distinct times.  In the latter case, we are likely to describe the situation as a stochastic process.  Initially there is a random variable $A$, with probability distribution $P(A)$.  Then, the system undergoes a stochastic evolution described by a stochastic matrix $\Gamma_{B|A}$, which transforms $A$ into another variable $B$, with probability distribution $P(B)$.  However, $P(A)$ and $\Gamma_{B|A}$ are just convenient summaries of a joint distribution $P(A,B)$, since $\Gamma_{B|A}$ is a matrix of transition probabilities, i.e. conditional probabilities. It is evident that any joint probability distribution $P(A,B)$ may in principle arise in this scenario and also in the case where the variables refer to distinct systems, so that one does not have to know the causal relations between the two variables in advance in order to know that a joint probability distribution is the correct mathematical object to use for describing their correlations.  

The analog of this in quantum theory would be to always describe correlations between systems described by algebras $\mathfrak{A}_A$ and $\mathfrak{A}_B$ by a joint state $\rho_{AB} \in \mathfrak{A}_A \otimes \mathfrak{A}_B$, regardless of whether $\mathfrak{A}_A$ and $\mathfrak{A}_B$ refer to two distinct systems or to the same system at two distinct times.  In the former case, this is indeed what we usually do.  However, in the latter case, we normally ascribe a state $\rho_A \in \mathfrak{A}_A$ to the system initially, and then assert that it evolves in time according to the TPCP map $\mathcal{E}_{B|A}: \mathfrak{A}_A \rightarrow \mathfrak{A}_B$ to obtain a state $\rho_B \in \mathfrak{A}_B$.  This is analogous to the stochastic process description given in the classical case above, but in the quantum case we do not normally associate this with a joint state $\rho_{AB}$.  The Choi-Jamio{\l}kowski isomorphism asserts that a description in terms of a joint state is indeed possible, since the map $\mathcal{E}_{B|A}$ is isomorphic to a conditional state $\rho_{B|A}$ from which a joint state $\rho_{AB}$ can be built by combining with $\rho_A$ via eq. (\ref{CDO:Reverse}).  Since we can also go in the other direction, we could equally well describe things just by specifying $\rho_{AB}$.  However, this is not quite enough, since we would like to assert that $\rho_{AB}$ provides an equally useful summary of the probabilistic predictions that may be obtained in this scenario, without having to go back and reconstruct $\rho_A$ and $\mathcal{E}_{B|A}$ via the isomorphism before calculating them.  In fact, this is almost the case, but a slight modification is needed and we actually consider the following series of correspondences:  
\begin{equation}
\left ( \rho_{AB} \right ) \leftrightarrow \left ( \rho_A, \rho_{B|A} \right ) \leftrightarrow \left ( \rho_A^T, \mathcal{E}_{B|A} \right ),
\end{equation}
where $^T$ denotes the transpose in the basis used to construct the isomorphism.  The transpose is related to a time reversal implicit in the construction, which is discussed in \cite{Leifer06}, but note that if an eigenbasis of $\rho_A$ is used to construct the isomorphism then $\rho_A^T = \rho_A$, so this would be a natural constraint to impose on the construction.

To understand how this works, it is helpful to return briefly to the classical case. If Alice has access to a random variable $A$ and Bob has access to a random variable $B$, and they ascribe the joint probability distribution $P(A,B)$ to the two variables, then the set of joint probability distributions they can generate by local processing of their variables is the same, regardless of whether Alice and Bob read $A$ and $B$ from distinct physical systems or if Bob's variable comes from the same physical system, sent to him through a noisy channel by Alice.  Essentially the same thing is true in the quantum case, although noncommutativity makes things a little more subtle.  In the case where $\rho_{AB}$ represents the joint state of two systems, the local processing consists of measurements on $\mathfrak{A}_A$ and on $\mathfrak{A}_B$.  However, in the case where it represents the same system at two different times, we have to move to a ``prepare and measure'' scenario where the local processing consists of a choice of an ensemble preparation for Alice and a measurement for Bob.  To describe this, we need to recall the formalism of generalized measurements in quantum theory.

A general measurement can be represented by a Positive Operator Valued Measure (POVM).  This is a collection of positive operators $\bm{M} = \{M_j\}$ in an algebra $\mathfrak{A}$ that sum to the identity $\sum_j M_j = I$.  The probability of obtaining the outcome $\bm{M}=j$ when the system is in state $\rho \in \mathfrak{A}$ is given by the generalized Born rule $\text{prob}(\bm{M}=j) = \Tr{M_j \rho}$.  It is less commonly appreciated that POVMs can also be used to describe ensemble preparation procedures.  This is demonstrated by the following lemma, which is proved in \cite{Leifer06}.\\

\noindent{\bf Lemma:}
Let $\rho$ be a state in $\mathfrak{A}$.  $\{p_j,\rho_j\}$ is an ensemble decomposition of $\rho$ iff there exists a POVM $\bm{M} = \{M_j\}$ such that $p_j = \Tr{M_j \rho}$ and $\rho_j = \frac{\sqrt{\rho} M_j \sqrt{\rho}}{\Tr{M_j \rho}}$.\\

Therefore, given a state $\rho$ and a POVM $\bm{M}$, there are two procedures that they could be used to describe.  An $\bm{M}$-\emph{measurement} of $\rho$ is a procedure that takes a system in the state $\rho$ as input and outputs a classical random variable with distribution $\text{prob}(\bm{M} = j) = \Tr{M_j \rho}$.  Conversely, an $\bm{M}$-\emph{preparation} of $\rho$ consists of first generating a classical random variable with distribution $\text{prob}(\bm{M} = j) = \Tr{M_j \rho}$ and then preparing the corresponding state $\rho_j = \frac{\sqrt{\rho} M_j \sqrt{\rho}}{\Tr{M_j \rho}}$ as output.  We are now in a position to state the main result, which is proved in \cite{Leifer06}.\\

\noindent{\bf Theorem:}
Let $\rho_{AB} \in \mathfrak{A}_A \otimes \mathfrak{A}_B$ be a state with reduced state $\rho_A \in \mathfrak{A}_A$ and conditional state $\rho_{B|A}$.  Let $\mathcal{E}_{B|A}$ be the TPCP map isomorphic to $\rho_{B|A}$ and let $^T$ denote the transpose of an operator taken in the basis used to define the isomorphism.  Let $\bm{N} = \{N_j\}$ be a POVM on $\mathfrak{A}_A$ and let $\bm{M} = \{M_k\}$ on $\mathfrak{A}_B$.  Then, the joint probability of getting outcome $j$ in an $\bm{N}$-measurement on $\mathfrak{A}_A$ and getting outcome $k$ in an $\bm{M}$-measurement on $\mathfrak{A}_B$, on a joint system in the state $\rho_{AB}$, is the same as the joint probability for obtaining the $j$ value of the classical input in an $N^T$-preparation of $\rho_A^T$ and getting outcome $k$ in an $\bm{M}$-measurement on $\mathfrak{A}_B$, when the system is evolved according to $\mathcal{E}_{B|A}$ between preparation and measurement.  Equivalently,
\begin{equation}
\label{CJI:MainEq}
\text{prob}(\bm{N} = j,\bm{M} = k) = \PTr{AB}{N_j \otimes M_k \rho_{AB}} = \PTr{B}{M_k \mathcal{E}_{B|A} \left ( \sqrt{\rho^T} N_j^T \sqrt{\rho^T} \right )}.
\end{equation}\\

\section{Subjectivity of Quantum Operations}

\label{Subjective}

We now turn to the question of what the above result means for the status of quantum operations in the CFS view of quantum theory.  The first point is that, when considering the probabilities of local measurements made on a bipartite system, the description we would normally give in terms of a bipartite state $\rho_{AB}$ can always be replaced by a description in terms of the pair $(\rho_A^T,\mathcal{E}_{B|A})$ via eq. (\ref{CJI:MainEq}).  The latter description looks just like a ``prepare and measure'' scenario, in which the TPCP map $\mathcal{E}_{B|A}$ describes the time evolution between preparation and measurement, even though we are ``actually'' talking about the correlations between two subsystems at a given time. In this context, CFS would assert that the assignment of the state $\rho_{AB}$ \emph{always} represents some agent's degree of belief and is \emph{never} to be thought of as representing an objective state of affairs.  Clearly, for this to be true, at least one of $\rho_A^T$ or $\mathcal{E}_{B|A}$ must represent degrees of belief rather than objective facts.  In fact, for CFS, \emph{both} $\rho_A^T$ and $\mathcal{E}_{B|A}$ must represent degrees of belief, because they impose no a priori constraints on the degree to which two agents' state assignments may differ, and both $\rho_A^T$ and $\mathcal{E}_{B|A}$ must be allowed to vary in order to obtain an arbitrary $\rho_{AB}$.   In particular, CFS state that $\rho_{AB}$ is subjective even if it is pure\footnote{The arguments for this will not be rehashed here, but see \cite{CFS06}}, and demanding purity of $\rho_{AB}$ is equivalent to demanding that $\mathcal{E}_{B|A}$ is an isometry, and unitary if $\mathfrak{A}_A = \mathfrak{A}_B$.  Thus, we already have a case where CFS would have to regard a unitary operation as representing subjective degrees of belief rather than an objective state of affairs.

However, in this case the unitary operation is simply providing part of a description of a bipartite system and the real question is whether unitary operations should be regarded as subjective when they are being used to describe the time evolution of a single system.  To argue this case, we introduce a variant of Leibniz's principle of the ``identity of indiscernibles'' \cite{SEPIdent}.  If the sum total of probabilistic assignments that can be made in one experimental scenario is identical to those that can be made in another scenario, then it is clear that both scenarios should be representable by an identical mathematical formalism.  Our principle states that we should ascribe subjectivity and objectivity to the elements of the formalism identically in both cases.  In the present context, if we are ``really'' using $(\rho_A^T,\mathcal{E}_{B|A})$ to describe a ``prepare and measure'' scenario, then $\mathcal{E}_{B|A}$ does represent a time evolution and the statistical predictions that can be made are identical to those of the bipartite scenario described above.  Thus, our principle requires that if $\mathcal{E}_{B|A}$ is subjective in the bipartite scenario, then it is also subjective when used to describe time evolution in the prepare and measure scenario.  In particular, for CFS, this has to apply to unitary operations, since they are treated as subjective in the bipartite scenario.

At this point, the subjectivity of unitary operations hangs on whether or not one accepts the principle described above.  The main argument for accepting it rests on the virtue of probabilistic abstraction, which is familiar in the classical case.  Consider the Kolmogorovian formulation of probability theory, in which we have a sample space of events.  This is a purely abstract mathematical theory and no identification between events in the sample space and physical events in spacetime is supposed.  Clearly, this is the reason behind the fact that we can describe spacelike and timelike correlated variables via an identical formalism, using joint probability distributions in both cases.  Now, subjectivist axiomatizations of probability theory, such as those provided by de Finetti and Savage \cite{deFinetti, Savage}, are focussed on deriving a mathematical representation of degrees of belief in various events from their \emph{logical} relations, rather than anything to do with how those events are embedded in spacetime.  This is natural for a theory which is about rational decision making in general, rather than being just about its application in physics, and leads to the abstraction of the theory from the details of causality.  If we are really to regard the better part of quantum theory as a ``law of thought'', as advocated by Fuchs \cite{Fuchs02, Fuchs03}, then it seems that we ought to adopt a similar approach as far as possible.  The fact that two scenarios entail the same set of possible probability ascriptions, is enough to guarantee their equivalence from the point of view of decision making.  Therefore, the two cases should be identified within the abstract theory, and the principle follows.

\section{Objectivity of Physical Laws}

\label{Reconcile}

Accepting the preceding argument implies that the Hamiltonians and Lagrangians of physics represent subjective degrees of belief, since assuming that we are prepared to regard time intervals as objective, the Hamiltonian of a system uniquely determines the unitary time evolution operator.  Our most fundamental physical theories, such as the standard model of particle physics, are essentially postulations of a particular Hamiltonian or Lagrangian, so it might seem that we are in danger of losing the objectivity of physical law altogether.

However, this worry is unfounded, and rests on a similar category error as the identification of objective certainty with probability one \cite{CFS06} (assuming a finite sample space to avoid the necessary caveats about sets of measure zero).  To the radical probabilist, these are very distinct assertions.  The statement that the probability of an event is equal to one is relative to the particular agent who asserts it.  It is verified by observing the agent's decision making behavior, e.g. asking her to enter into a bet on the event and finding out that she is willing to bet her life on it.  On the other hand, objective certainty means that the event is sure to occur and can only be verified by empirical observation of the occurrence of the event itself, or by logical deduction from other objective certainties.  For the radical probabilist, agents may make probability one assignments even if the event itself is not an objective certainty because no prior probability assignment is ruled out as irrational a priori.  To be sure, a probability one assertion entails a rather strong commitment on the part of the agent, and it does mean that the agent \emph{believes} that the event is certain to occur.  In particular, if she believes that it is an objective certainty then she must assign probability one.

The fact that probability one is not identified with objective certainty does not mean that objective facts about the world do not exist, just that they have no representation in probability theory without reference to an agent who believes in them.  Similarly, if Hamiltonians are taken as subjective degrees of belief rather than objective physical laws it does not mean that objective physical laws have no bearing on Hamiltonian assignments.  Belief in the truth of a particular physical law, can indeed constrain the class of Hamiltonians that an agent may assign.  For example, if the Hamiltonian does not respect a particular symmetry principle, such as Lorentz invariance, that the agent believes to be true then it is not a legitimate representative of the agent's beliefs.  Here it is the symmetry principle, and not the Hamiltonian itself, that captures the objective content of the physical law.

\section{Conclusions}

\label{Conc}

To summarize, we have argued that if quantum states, including pure states, are to be regarded as representing subjective degrees of belief, then it is natural to regard quantum operations, including unitary ones, as also being subjective. Essentially, if quantum states, including pure states, are more like probability distributions than ``states of reality'', then quantum operations, including unitary ones, are more like conditional probabilities than objective dynamical laws and should likewise be taken to be subjective.

Perhaps more importantly, this work raises the question of whether a formalism for quantum theory could be given that does not require causal relations to be specified a priori.  Although quantum theory is often thought to be a kind of generalized probability theory, it is not often formulated at the same level of abstraction as the classical theory.  In the usual formulation of quantum theory, when we speak of joint states we are referring to the state of two distinct subsystems and when we speak of correlations between the same system at two different times we use TPCP maps instead.  As noted above, this is a closer analog to the classical theory of stochastic processes than it is to a fully abstract probability theory.  In the canonical framework for quantum theory, this same issue is manifested in the fact that quantum states are always referred to spakelike hypersurfaces rather than to arbitrary collections of regions in spacetime.  In other words, we need to know some minimal information about the causal relations between events before we can even set up the theory.  We take the current work as a demonstration that, in fact, joint quantum states need not be exclusively referred to spacelike separated regions, but can also be used to describe the correlations between algebras referring to potentially timelike separated events.  This indicates that it may be possible to formulate quantum theory at the same level of abstraction as Kolmogorov probability theory, although much further work is needed to realize this possibility. Having such a formalism would hopefully shed further light on the foundations of quantum theory and quantum information, and may even play a role in the construction of a background independent quantum theory of gravity, wherein there is good reason to suspect that causal relations between events may not be fixed a priori.\\

\noindent{\bf Acknowledgments:}  I would like to thank Rob Spekkens for many interesting discussions on topics related to this work.  I would also like to thank Andrei Khrennikov and Chris Fuchs for inviting me to speak at this conference, and for allowing me to fill this article with speculations that may or may not prove meaningful.  Research at Perimeter Institute is supported in part by the Government of Canada through NSERC and by the Province of Ontario through MEDT.  At Cambridge, this work was supported by the European Commission through QAP, QAP IST-3-015848, and through the FP6-FET Integrated Project SCALA, CT-015714.\\

\bibliographystyle{plain}
\bibliography{JamBib}

\end{document}